\documentclass{camera}
\usepackage{graphicx}  

\usepackage{bm}
\usepackage{color}

\usepackage{amssymb}

\begin{document}

%
\title{Theoretical studies of possible toroidal high-spin isomers in the light-mass region}

%
\vspace*{-0.6cm}
\author{Andrzej Staszczak$^1$, Cheuk-Yin~Wong$^2$}

%
\vspace*{-0.6cm}
\organization{
$^1$Institute of Physics, Maria Curie-Sk{\l}odowska University, Lublin, Poland\\
$^2$Physics Division, Oak Ridge National Laboratory, Oak~Ridge, TN USA
}

\maketitle
\vspace*{-0.6cm}
\begin{abstract}
We review our theoretical knowledge of possible toroidal high-spin isomers
in the light mass region in 28$\le$$A$$\le$52 obtained previously in cranked
Skyrme-Hartree-Fock calculations. We report additional toroidal high-spin isomers
in $^{56}$Ni with $I$=114$\hbar$ and 140$\hbar$, which follow the same
(multi-particle)--(multi-hole) systematics as other toroidal high-spin isomers.
We examine the production of these exotic nuclei by fusion of various projectiles on
$^{20}$Ne or $^{28}$Si as an active target in time-projection-chamber (TCP) experiments.
\end{abstract}

%
\vspace*{-0.6cm}
\section{Introduction}
Wheeler suggested that under appropriate conditions the nuclear
fluid may assume a toroidal shape \cite{Gam61}. If toroidal nuclei
could be made, there would be a new family tree for the investigation
of the nuclear species.

The rotating liquid-drop model is useful as a qualitative guide
to point out essential energy balances leading to possible toroidal
figures of equilibrium \cite{Won73}. A quantitative assessment relies on
microscopic descriptions that include both the bulk properties of the
nucleus and the single-particle shell effects in self-consistent
mean-field theories, such as the Skyrme-Hartree-Fock (SHF) approach
\cite{Vau72}. The non-collective rotation with an angular momentum
about the symmetry axis is permissible quantum mechanically for an axially
symmetric toroid by making particle-hole excitations and aligning
the angular momenta of the constituents along the symmetry axis \cite{Boh81}.
As a consequence, only a discrete, quantized set of total angular momentum
$I$=$I_z$ states is allowed.

In our recent works \cite{Sta14}, we showed by using a cranked SHF approach
that toroidal high-spin isomeric states have general occurrences in 28$\leq$$A$$\leq$52
for even-even $N$=$Z$ and $N$$\ne$$Z$ nuclei. Toroidal high-spin isomers
have also been found theoretically in similar HF calculations in this mass
region \cite{Ich12}.

We would like to review the systematics of these  nuclei and suggest ways
how these nuclei may be produced.

\section{Light-mass toroidal high-spin isomers}

\begin{figure}[htb]
\begin{center}
\includegraphics[width=0.65\columnwidth]{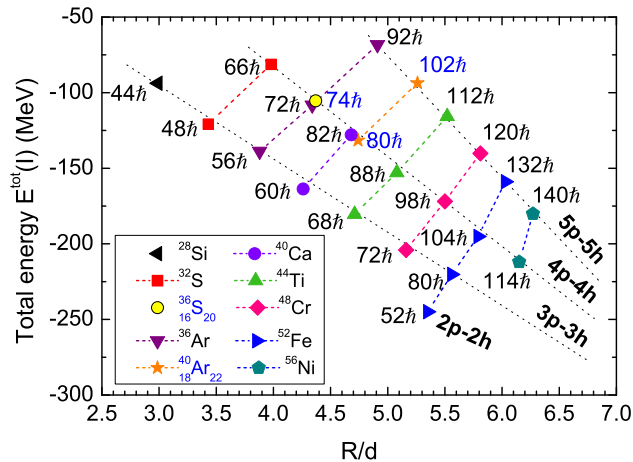}
  \caption{\label{Fig1} (Color online.) The total energies of the
    isomeric toroidal states of ${}_{14}^{28}$Si, ${}_{16}^{32}$S,
    ${}_{16}^{36}$S, ${}_{18}^{36}$Ar, ${}_{18}^{40}$Ar,
    ${}_{20}^{40}$Ca, ${}_{22}^{44}$Ti, ${}_{24}^{48}$Cr, ${}_{26}^{52}$Fe,
    and ${}_{28}^{56}$Ni and their associated angular momenta $I$=$I_z$ values along the symmetry axis,
    as a function of $R/d$. The $n$p-$n$h configurations relative to the $I$=0
    states are also indicated.}
\end{center}
\end{figure}

We have located the toroidal high-spin isomers at their energy minima using
the cranked SHF approach \cite{Sta14}. We find that the ratio of the torus major
radius $R$ to the torus minor radius $d$, $R/d$, increases with angular momentum
and approximately linearly with the mass number while the minor radius $d$
remains essentially unchanged (see Ref.~\cite{Sta14}). It is useful to classify
these nuclei according to the $n$-particle $n$-hole nature of the isomer,
relative to the toroidal nucleus at $I=0$. One finds that all $n$p-$n$h
families follow a regular well-behaved pattern as shown in Fig.~\ref{Fig1},
where we plot the total energy $E^{\rm tot}(I)$ of the isomer $^A  Z^t$($I$)
as a function of $R/d$ for different toroidal isomers with various aligned
angular momenta $I$.

\vspace*{-0.3cm}
We collect  the  properties of all known 21 toroidal high-spin isomers up to
$^{52}$Fe obtained previously in \cite{Sta14} in Fig.~\ref{Fig1}.
These systematics predict the possible presence of $n$p-$n$h toroidal high-spin
isomers for $^{56}$Ni at $R/d$ $\sim$ 6.0. Indeed, we found energy minima for
$^{56}$Ni with $I$=114$\hbar$ and 140$\hbar$ at $R/d$=6.15 and  6.27, respectively,
in subsequent cranked SHF calculations.

\begin{figure}[htb]
\begin{center}
\includegraphics[width=0.7\columnwidth]{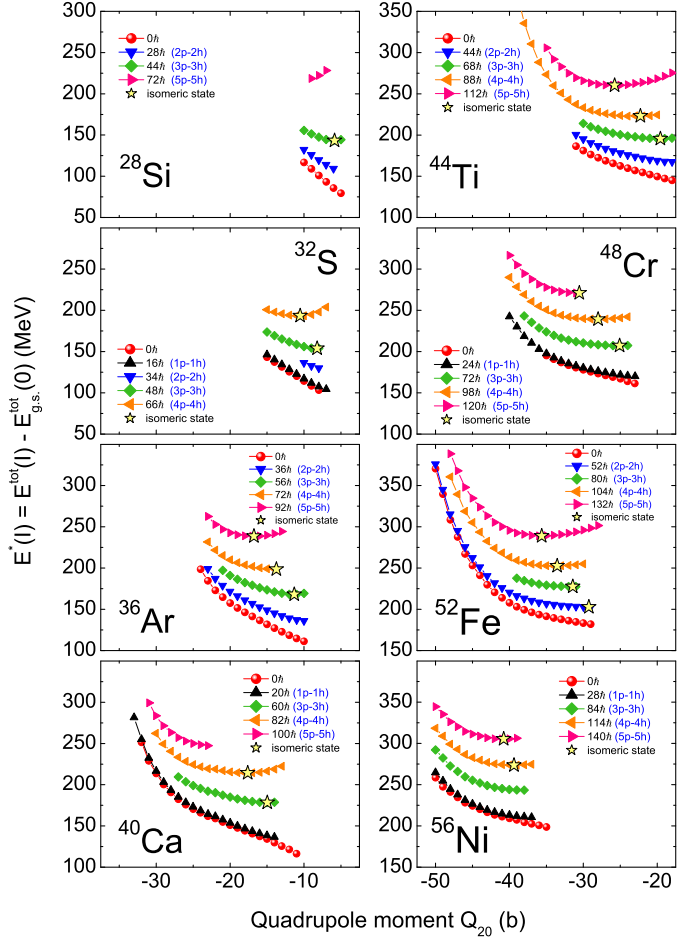}
  \caption{\label{Fig2} (Color online.) The excitation energies $E^*(I)$  of high-spin
    toroidal states  of $^{28}$Si, $^{32}$S, $^{36}$Ar, $^{40}$Ca,
    $^{44}$Ti, $^{48}$Cr, $^{52}$Fe, and $^{56}$Ni as a function of the quadrupole moment
    $Q_{20}$ for different angular momenta along the symmetry axis, $I$=$I_z$.}
\end{center}
\end{figure}

\vspace*{-0.3cm}
\noindent
In Fig.~\ref{Fig2} the excitation energies $E^*(I)$ of even-even $N$=$Z$ toroidal states
relative to the energy of the ground states  are presented as a function of the quadrupole
moment $Q_{20}$ for different $I$ values along the symmetry axis. They include those from
our earlier work in \cite{Sta14} and the newly found $^{56}$Ni toroidal  isomers.
The locations of energy minima  $E^*({}^AZ^t(I))$ representing toroidal high-spin
isomeric states are indicated by star symbols.

\section{Production of light-mass toroidal isomers }

As the question of the mean half-life of these isomers remains unresolved, one can design
experiments that could detect the existence of the exotic toroidal high-spin nuclei
of Figs.~\ref{Fig1} and \ref{Fig2} at appropriate energies, if the mean half-lives
are longer than $\hbar$/MeV  or 200 fm/c. One way is to search for these isomers
as resonances or metastable nuclei by bombarding projectile nucleus ${}^{A_p} Z_p$
on an active-target nucleus ${}^{A_T} Z_T$ for the production of the toroidal high-spin
isomer ${}^{A} Z^t(I)$ with angular momentum $I$=$I_z$,
\begin{eqnarray}
{}^{A_p} Z_p + {}^{A_T}Z_T  \to {}^{A}  {Z^t}(I).
\label{eq1}
\end{eqnarray}
The active target can be, for example, $^{20}$Ne, $^{36,38,40}$Ar, or $^{28}$Si.
We shall consider the cases of $^{20}$Ne and $^{28}$Si as active-targets.

In recent years, TPC chambers have been used to study the nuclear spectroscopy of
metastable nuclei \cite{Ahn12,Rog11}. The idea is to use a chamber of noble gas under
a high voltage so that the gas itself or an embedded solid layer serves as the target,
and the nuclear trajectories show up as tracks. The production of a composite nucleus
with a long half-life would show up as a single track with the mass and charge arising
from the fusion of the projectile and target nuclei.
The production of binary products indicates a two$\to$two reaction from which one
can examine the elastic and inelastic channels and study the excitation function and
angular distribution to search for various meta-stable states. Previously, many metastable
states formed by colliding various projectile nuclei with an active He target have been
found by such a technique \cite{Ahn12}.

The cross section for producing a toroidal isomer at the correct energy and angular momentum
is \cite{PDG14} (p. 517)
\begin{eqnarray}
\sigma_{\rm res}(E, {}^{A}{Z}^t(I))=\frac{4\pi}{k^2} (2I+1)
\frac{\Gamma^2/4}{[E-E_{\rm res}({}^A {Z}^t(I))]^2+\Gamma^2/4}B_{\rm in} B_{\rm out},
\end{eqnarray}
where $E$ is the c.m. energy, $E_{\rm res}({}^A {Z}^t(I))$ is the c.m. resonance energy
for the toroidal high-spin isomer with spin $I$, $k$ is the c.m. momentum in the initial
state, and $\Gamma$ is the full width at half maximum height of the resonance.
The quantities $B_{\rm in}$ and $B_{\rm out}$ are the branching fractions for the resonance
into the initial-state and  finial-state channel, respectively.
Here, the width $\Gamma$ and the branching fractions $B_{\rm in}$ and $B_{\rm out}$ may
need to be determined experimentally. The resonance energy $E_{\rm res}$ (in the c.m. system)
is given by
\begin{eqnarray}
E_{\rm res}( {}^{A} Z^t(I))  =M({}^{A} Z^t(I))- M({}^{A_T}Z_T ) -M({}^{A_p} Z_p),
\label{eq4}
\end{eqnarray}
where  $M(^{A} Z^t(I))$, $M(^{A_T}Z_T )$, and $M(^{A_p} Z_p)$ are nuclear masses of
$^A Z^t$($I)$, $^{A_T}Z_T$, and $^{A_p} Z_p$, respectively. In terms of the
binding energies $B({}^{A} Z)$, $ B({}^{A_T} Z_T )$, and $B({}^{A_p} Z_p)$,
and excitation energy $E^*({}^A Z^t(I))$, we have
\begin{eqnarray}
E_{\rm res}( {}^{A} Z^t(I))
= E^*({}^A Z^t(I))- B({}^{A} Z) + B({}^{A_T} Z_T )  + B({}^{A_p} Z_p).
\end{eqnarray}

\vspace{0.5cm}
\begin{figure}[htb]
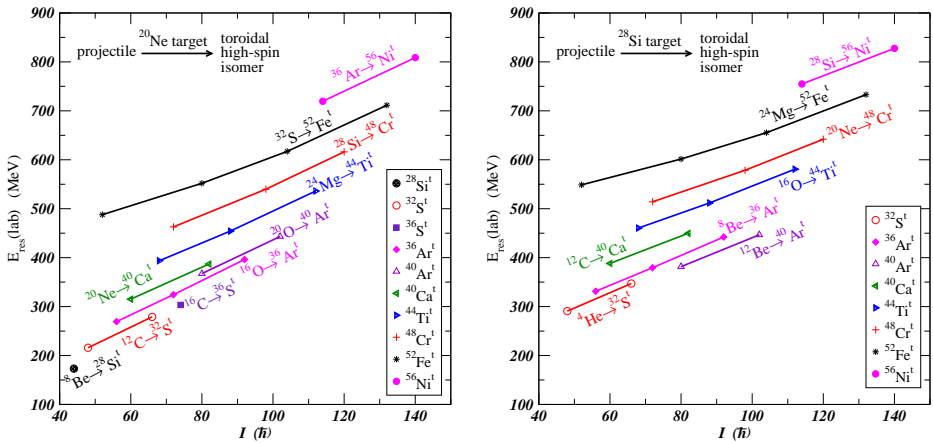

\begin{minipage}[b]{0.5\linewidth}
\centering
\includegraphics[width=0.93\columnwidth]{fig3a}
\end{minipage}%
\begin{minipage}[b]{0.5\linewidth}
\centering
\includegraphics[width=0.93\columnwidth]{fig3b}
\end{minipage}%
\caption{\label{Fig3} (Color online) The resonance energies $E_{\rm res}( {}^{A} Z^t(I))({\rm lab})$
  in the laboratory system in the bombardment of various projectiles $^{A_p}Z_p$ on $^{20}$Ne (left panel)
  or $^{28}$Si (right panel) as an active target for the production of toroidal isomers $^{A}Z^t$($I$)
  with different angular momenta $I$.
}
\end{figure}

\noindent
The resonance energy $E_{\rm res}( {}^{A} Z^t(I))({\rm lab})$ in the laboratory system
is given by $E_{\rm res}( {}^{A} Z^t(I))\times (A_p+A_T)/A_T$. For the production of
toroidal high-spin isomers by collision on $^{20}$Ne or $^{28}$Si as an active target,
the resonance energies are shown in Fig.~\ref{Fig3}. The knowledge of the predicted
values of $E_{\rm res}( {}^{A} Z^t(I))({\rm lab})$ will facilitate the search of toroidal
high-spin isomers.

\vspace*{-0.3cm}
\section{Conclusions and discussion}

Under (multi-particle)--(multi-hole) excitation involving large orbital angular momentum
orbitals,  non-collective rotations of many  light nuclei lead to equilibrium configurations
whose densities may assume the shape of a torus. The $n$p-$n$h systematics of toroidal
high-spin isomers fit a regular pattern which can be used to predict possible presence
of toroidal high-spin isomer in $^{56}$Ni. We found additional equilibrium energy minima
for $^{56}$Ni with $I$=114$\hbar$ and 140$\hbar$ in subsequent cranked SHF calculations.

We examine the production of  light-mass toroidal high-spin isomers by fusion of various
projectile nuclei with an active target, in which the trajectories of the reaction products
can be examined as a function of the collision energies. Resonance energies for the production
of toroidal high-spin isomers have been calculated for $^{20}$Ne or $^{28}$Si as an  active
target, based on the Skyrme-Hartree-Fock energies obtained for the isomers.

The technology of building a TCP using 90\% of $^{20}$Ne as its dominant ingredient has been
developed by the ALICE collaboration \cite{Ket13}. The utilization of a similar TPC detector
with $^{20}$Ne or $^{28}$Si as an active target for nuclear spectroscopy may prove useful
for the search of toroidal isomers.

\vspace*{-0.3cm}
\section*{Acknowledgements}
The authors wish to thank Profs. J. Natowitz and K. Read for helpful discussions.
This work was supported in part by the Division of Nuclear
Physics, U.S. Department of Energy, Contract No. DE-AC05-00OR22725.



%

\vspace*{-0.3cm}
\begin{thebibliography}{99}

\bibitem{Gam61}
See a reference to J. A. Wheeler's toroidal nucleus in
Gamow G. 1961 {\it Biography of Physics}
(New York: Harper \& Brothers  Publishers) pp.~297.

\bibitem{Won73}
Wong C.Y.,
{\it Ann. Phys.} {\bf 77} (1973) 279;
{\it Phys. Rev. C} {\bf 17} (1978) 331.

\bibitem{Vau72}
Vautherin D. and Brink D.M.,
{\it Phys. Rev. C} {\bf 5} (1972) 626;
Engel Y.M., Brink D.M., Goeke K., Krieger S., and Vautherin D.,
{\it Nucl. Phys.} {\bf A 249} (1975) 215.

\bibitem{Boh81}
Bohr A. and Mottelson B.R.,
{\it Nucl. Phys.} {\bf A 354} (1981) 303c.

\bibitem{Sta14}
Staszczak A. and Wong C.Y.,
{\it Phys. Lett. B} {\bf738} (2014) 401;
{\it Acta Phys. Pol. B} {\bf 46} (2015) 675, arXiv:1504.07646;
{\it Phys. Scripta} (2015) in press, arXiv:1412.0050.

\bibitem{Ich12}
Ichikawa T., Maruhn J.A., Itagaki N., Matsuyanagi K., Reinhard P.-G., Ohkubo S.,
{\it Phys. Rev. Lett.} {\bf 109} (2012) 232503;
Ichikawa T., Matsuyanagi K., Maruhn J.A., and Itagaki N.,
{\it Phys. Rev. C} {\bf 89} (2014) 011305(R);
{\it Phys. Rev. C} {\bf 90} (2014) 034314.

\bibitem{Ahn12}
Suzuki D. {\it et al.},
{\it Nucl. Instr. Meth. A} {\bf 691} (2012) 39;
Suzuki D. {\it et al.},
{\it Phys. Rev. C} {\bf 87} (2013) 054301.

\bibitem{Rog11}
Roger T. {\it et al.},
{\it Nucl. Instr. Meth. A} {\bf 638} (2011) 134, arXiv:1012.3560.


\bibitem{PDG14}
Olive K.A. {\it et al.},
(Particle Data Group),
{\it Chin. Phys. C} {\bf 38} (2014) 1.

\bibitem{Ket13}
B. Ketzer for the GEM-TPC, ALICE TPC Collaborations,
{\it Nucl. Instr. Meth. A} {\bf 732} (2013) 237, arxiv:1303.6694.

\end{thebibliography}
\end{document}